# Probing the Chiral Anomaly by Planar Hall Effect in Three-dimensional Dirac Semimetal $Cd_3As_2$ Nanoplates

Min Wu[1,2,*], Guolin Zheng[1,*], Weiwei Chu[1,2], Wenshuai Gao[1,2], Hongwei Zhang[1,2], Jianwei Lu[1,2], Yuyan Han[1], Jiyong Yang[1], Haifeng Du[1,3,4], Wei Ning[1,§], Yuheng Zhang[1,2,4] and Mingliang Tian[1,2,3,4,§]

[1]*Anhui Province Key Laboratory of Condensed Matter Physics at Extreme Conditions, High Magnetic Field Laboratory, the Chinese Academy of Science (CAS), Hefei 230031, Anhui, the People's Republic of China.*

[2]*Department of physics, University of Science and Technology of China. Hefei 230026, the People's Republic of China.*

[3]*Department of Physics, College of Physics and Materials Science, Anhui University, Hefei 230601, Anhui, the People's Republic of China.*

[4] *Collaborative Innovation Center of Advanced Microstructures, Nanjing University, Nanjing 210093, the People's Republic of China.*

[*] Those authors contribute equally to this work.

[§]To whom correspondence should be addressed. E-mail: ningwei@hmfl.ac.cn (W. N.), tianml@hmfl.ac.cn (M.T.).

# Abstract

Searching for exotic transport properties in new topological state of matter is an active topic. One of the most fascinating achievements is the chiral anomaly in recently discovered Weyl semimetals (WSMs), which is manifested as a negative longitudinal magnetoresistance (LMR) in the presence of a magnetic field ***B*** parallel to an electric field ***E***. Another predicted key effect closely related to the chiral anomaly is the planar Hall effect (PHE), which has not been identified in WSMs so far. Here we carried out the planar Hall measurements on $Cd_3As_2$ nanoplates, and found that, accompanied by the large negative LMR, a PHE with non-zero transverse voltage can be developed while tilting the in-plane magnetic field ***B*** away from the electric field ***E***. Further experiments reveal that both the PHE and the negative LMR can be suppressed synchronously by increasing the temperature, but still visible at room temperature, indicating the same origin of these two effects. The observation of PHE in $Cd_3As_2$ nanoplates gives another key transport evidence for the chiral anomaly, which may provide us with deep insight into the chiral charge pumping in Weyl Fermions system.

Weyl semimetal (WSM) is a new realization of topological state of matter, whose emergence has inspired intense interest in condensed matter physics, due to its exotic transport phenomena. One of the most exciting transport properties in WSMs is the chiral anomaly [1, 2], which refers to the non-conservation of chiral charges in the presence of a magnetic field parallel to an electric field ($\boldsymbol{E} \cdot \boldsymbol{B} \neq 0$) and is usually manifested as a large negative longitudinal magnetoresistance (LMR) [3-7]. If the in-plane magnetic field is not strictly parallel or perpendicular to the electric field ($\boldsymbol{E} \cdot \boldsymbol{B} = EB\cos\theta, \theta \neq 0, \pi/2$), a large non-zero transverse voltage can be developed, leading to a secondary effect closely related to the chiral magnetic effect-*Planar Hall effect* (PHE) [8, 9]. Theoretically, the PHE in non-magnetic WSMs can be strictly retrospected to the chiral magnetic effect, which provides another key transport signature of chiral anomaly in WSMs.

Previously, the negative LMR was usually labeled as the transport signature of chiral magnetic effect and has been observed in Dirac semimetals $Na_3Bi$ [10], $Cd_3As_2$ [11, 12], $ZrTe_5$ [13-15], GdPtBi [16] and Weyl semimetal TaAs family [17-19]. While recent studies indicated that the negative LMR can also be induced by the axial anomaly due to neutral and ionic impurities, inhomogeneous current distribution (or the "current jetting" effect) as well as conductivity fluctuations [20-23]. Thus, the negative LMR itself becomes an ambiguous transport evidence for the chiral anomaly. A "smoking gun" signature of chiral anomaly should be the combination of both the angular-dependent negative LMR and the PHE [8]. Unfortunately, the PHE in Dirac/Weyl semimetals has not been observed yet.

In this letter, we report the observation of PHE in Dirac semimetal $Cd_3As_2$ nanoplates. These nanoplates exhibit large negative LMR in **B** ∥ **E** configuration in low temperature region. Tilting the in-plane magnetic field **B** away from the electric field **E**, a large transverse voltage is detected. After subtracting the magnetoresistance components originating from the experimental misalignments, the obtained planar transverse magnetoresistance (TMR) displays a $\sim\cos\theta\sin\theta$ dependence, which is consistent with the PHE. Further study demonstrates that both the negative LMR and the PHE can be suppressed synchronously by increasing the temperature $T$, but they are still visible at room temperature ($T = 300$ K). The field-dependent amplitudes of the PHE and its correlation to the scattering time are also discussed.

$Cd_3As_2$ nanoplates were grown by chemical vapor deposition (CVD) method. The electrical contacts of the sample were fabricated by the standard electron-beam lithography (EBL) technique followed by Au (100 nm)/Ti (10 nm) evaporation and lift-off process. Fig. 1(a) shows the standard six-probe Hall-bar device prepared by EBL, the magnetic field $B$ is rotated in the sample plane with an angle $\theta$ to the current direction. The sample thickness is determined to be about $90\ nm$ by the 52°-tilted scanning electron microscopy (SEM), as shown in inset of Fig. 1(a). For tilted angle $\theta = 0°$, a large negative LMR can be identified with pronounced Shubnikov-de Haas (SdH) oscillations at 2 K, as shown in Fig. 1(b). By tilting the angle $\theta$ away from 0°, the negative LMR is suppressed gradually and it becomes positive above 40°. The peak positions of SdH oscillations are nearly unchanged, indicating an isotropic Fermi surface in $Cd_3As_2$. This angular-dependent negative

LMR is usually attributed to the chiral magnetic effect induced by the chiral charge imbalance in $\boldsymbol{B} \parallel \boldsymbol{E}$ configuration. Rotating the in-plane magnetic field from $0°$ to $360°$ consecutively at a fixed field magnitude ($|B| = 14\,\text{T}$), we get a highly anisotropic planar LMR $R_{xx}$, as shown in Fig. 1(c). As discussed above, $Cd_3As_2$ has an isotropic Fermi surface, this anisotropy of $R_{xx}$ can be generally attributed to the chiral magnetic effect. Theoretically, the planar LMR $R_{xx}$ should exhibit a $\sim\cos^2\theta$ dependence [8], this is qualitatively consistent with our observation in Fig. 1(c). While experimentally, the magnetic field $B$ cannot exactly locate in the sample plane (with the misalignment $|\delta\gamma| \leq 1°$) during the in-plane rotation. What's more, this out-of-plane component of field $B$ is usually $\theta$-dependent [24]:

$$\delta B_\perp = B \sin(\delta\gamma) \cos\theta, \tag{1}$$

leading to an ordinary orbital magnetoresistance component $\Delta R_{xx}^\perp$ which also exhibits a $\sim\cos^2\theta$ tendency. Moreover, this ordinary orbital magnetoresistance component $\Delta R_{xx}^\perp$ could be considerably large comparing to the intrinsic planar LMR, even for a small misalignment ($\leq 1°$), due to the extremely large orbital magnetoresistance in topological semimetals. Thus the intrinsic planar LMR $R_{xx}$ would be contaminated by $\Delta R_{xx}^\perp$, leading to an experimentally undistinguishable planar LMR. In the following, we will mainly discuss the planar TMR $R_{xy}$, which provides another transport signature of the chiral anomaly in WSMs.

For planar TMR $R_{xy}$, there are two types of misalignments. The first type of misalignment originates from the magnetic field misalignment, or the ordinary Hall resistance component induced by the out-of-plane field component $\delta B_\perp$, as discussed above in equation (1). The second type of misalignment comes from the transverse Hall bar misalignment during the nanofabrication, which leads to a small planar LMR component $\Delta R_{xx}^{\parallel}$ coupled to the planar TMR. Fig. 1(d) presents the planar TMR $R_{xy}$ versus $\theta$ with $B = \pm 14$ T, at $T = 2$ K and $250$ K, respectively. Theoretically, the PHE can be expressed as [8]:

$$\rho_{xy}^{PHE} = -\Delta\rho \sin\theta \cos\theta , \qquad (2)$$

where $\Delta\rho = \rho_{\perp} - \rho_{\parallel}$ is the resistivity anisotropy induced by chiral anomaly and should be positive, $\rho_{\perp}$ and $\rho_{\parallel}$ can be regarded as the resistivity corresponding to the current flowing perpendicular and parallel to the direction of magnetic field, respectively. At $T = 2$ K, the ordinary Hall resistance component induced by $\delta B_{\perp}$ would be considerably large, due to the extremely large Hall resistance (or the ultra-low carrier density) in our $Cd_3As_2$ nanoplates, and will largely overwhelm the intrinsic signals of PHE, making the PHE invisible, as we can see in Fig. 1(d). However, as $T$ increases the ordinary Hall resistance component can be significantly suppressed, making the PHE visible on $R_{xy}$ at high temperature. This can be seen in Fig. 1(d) where the extra kinks near $\theta = 45°$ and $135°$ on the planar TMR $R_{xy}$ can be easily identified at $T = 250$ K, which originates from the intrinsic PHE in $Cd_3As_2$ nanoplates. Since the PHE is not generated by the conventional Lorentz force, the planar TMR $R_{xy}$ does not satisfy the anti-symmetry property, that is to say $R_{xy} \neq -R_{yx}$. On the other hand, the ordinary Hall resistance induced by $\delta B_{\perp}$ is anti-symmetric with respect to B, thus the first kind of misalignment can in principle be subtracted by symmetrizing the planar TMR via the equation

$$R_{xy}^{sym} = [R_{xy}(B,\theta) + R_{xy}(-B,\theta)]/2. \tag{3}$$

Now we symmetrize the planar TMR $R_{xy}$ at $T = 2$ K to subtract the first type of misalignment. Fig. 2(a) shows the symmetrized magnetoresistance $R_{xy}^{sym}$ (blue squares) versus $\theta$ at a fixed field $|B| = 14$ T. It is clearly found that the $R_{xy}^{sym}$ does not display a $-\sin\theta\cos\theta$ tendency exactly, as described in equation (2). This is due to the second type of misalignment, the Hall bar misalignment during the nanofabrication, which leads to a small planar LMR component $\Delta R_{xx}^{\parallel}$ coupled to the $R_{xy}^{sym}$. As we have discussed above, the planar LMR exhibits a $\cos^2\theta$ tendency, thus the small planar LMR component $\Delta R_{xx}^{\parallel}$ can be subtracted by fitting the $R_{xy}^{sym}$ using the following formula:

$$R_{xy}^{Sym} = R_{xy}^{PHE} + \Delta R_{xx}^{\parallel}. \tag{4}$$

Where $R_{xy}^{PHE} = -a\sin\theta\cos\theta$, $\Delta R_{xx}^{\parallel} = b\cos^2\theta + c$. The parameter $a$ and $b$ are the amplitudes of PHE and $\Delta R_{xx}^{\parallel}$, respectively. The constant $c$ can be determined by the zero-field TMR, $R_{xy}$ ($B = 0$ T), which is about $-10$ Ω, as shown in inset of Fig. 2(a). Indeed, the equation (4) can fit well with $R_{xy}^{Sym}$, as we can see the red curve in Fig. 2(a). After subtracting the planar LMR component $\Delta R_{xx}^{\parallel}$, we finally get the intrinsic PHE shown in Fig. 2(b).

Such a large intrinsic PHE in Dirac/Weyl semimetals has not been identified thus far. Previously, PHE was only observed in ferromagnetic materials and was usually explained by the in-plane ferromagnetic order combined with spin-orbit couplings [25, 26]. The large PHE observed here in non-magnetic Dirac semimetal $Cd_3As_2$ nanoplates is related to the nontrivial Berry curvatures that couple the external magnetic field ***B*** to the collinear electric field ***E***, leading to the chiral charge imbalance. That is to say, PHE strictly originates from the chiral anomaly, which can provide another key transport signature of chiral magnetic effect.

Since the PHE can be strictly retrospected to the chiral anomaly, the PHE should exhibit a similar temperature-dependence as the negative LMR. That is, they should be suppressed synchronously as increasing the $T$. This can be testified by tracking both the PHE and the negative LMR under various temperatures, as shown in Fig. 3(a) and 3(b). Fig. 3(a) shows the temperature-dependent PHE ($R_{xy}^{PHE}$) at a fixed field $|B| = 14$ T. For $T < 250$ K, the amplitudes of $R_{xy}^{PHE}$ does not exhibit any attenuation as $T$ increased, while for $T \geq 250$ K, it decreases sharply. Similar tendency can also be found for the negative LMR in Fig. 3(b). As we can see, the negative LMR keeps unchanged below 250 K, but it is suppressed dramatically while increasing $T$ above 250 K. Such a synchronous behavior of both the PHE and the negative LMR with respect to $T$ reveals the same underlying physical origin of both effects: the chiral anomaly induced by the collinear magnetic field and electric field. It is worth noting that, both the PHE and the negative LMR persist up to $T = 300$ K, indicating that both effects are robust against thermal fluctuation.

For a deep insight into the observed PHE in Cd$_3$As$_2$ nanoplates, we now discuss the field-dependent amplitudes of $R_{xy}^{PHE}$. We have tracked the $R_{xy}^{PHE}$ under different magnetic field at $T = 2$ K and 100 K, as shown in Fig. 4(a) and 4(b), respectively. The amplitudes of PHE versus $B$ are shown in Fig. 4(c). As discussed above, the amplitudes of PHE below 100 K are nearly unchanged, this is consistent with the observed negative LMR in Fig. 3(b), where the chiral magnetic effect is not sensitive to the temperature below 100 K. Fitting the amplitudes of PHE to a $B^\alpha$-type function generates a parameter $\alpha \approx 2.0$ for both $T = 2\,\text{K}$ and $100\,\text{K}$. According to the equation (2), as systemically discussed in Ref. 8, the amplitudes of PHE are determined by the resistivity anisotropy induced by the chiral anomaly. For the weak magnetic field limit, $R_{xy}^{PHE}$ has the following relationship [8],

$$R_{xy}^{PHE} \propto (L_c/L_a)^2 \sin\theta \cos\theta. \tag{5}$$

Where $L_a = D/\Gamma B$ is the field-related length scale with $D$ the diffusion coefficient and $\Gamma$ the transport coefficient. $L_c = \sqrt{D\tau_c}$ is the chiral charge diffusion length with $\tau_c$ the scattering time of chiral charges. While for strong field limit ($L_a \ll L_c$) or under the quantum limit (only the lowest Landau level is occupied), the amplitudes of PHE are sample size dependent and tend to be saturated. It seems that the observed PHE satisfies the weak magnetic field limit (or $L_a \gg L_c$), which exhibits a $B^2$ tendency. However, it is noted that, the SdH oscillation frequency in Fig. 1(b) is about 10 T. That is to say, the Fermi level can be driven into quantum limit above 10 T, which is actually beyond the weak field limit while above 10 T. The $B^2$-dependence of the amplitudes of the PHE suggests that the scattering time $\tau_c$ is probably small in our nanoplates, leading to the criteria of weak field limit $(L_a \gg L_c)$. This assignment is also supported by the fact that the observed negative LMR does not exhibit a sharp decreasing in low field region and also the negative LMR does not saturate even with $B$ up to 14 T, which indicates a small scattering time $\tau_c$. We believe that the small scattering time $\tau_c$ is probably due to the relative high Fermi level and large carrier density in our $Cd_3As_2$ nanoplates [4-5, 12].

To summarize, we studied the planar Hall effect in Dirac semimetal $Cd_3As_2$ nanoplates. Besides the large negative LMR in the presence of a magnetic field ***B*** parallel to an electric field ***E***, we clearly demonstrated a non-zero transverse magnetoresistance at low temperatures while tilting the in-plane magnetic field ***B*** away from the electric field ***E***. This non-zero transverse magnetoresistance is attributed to the chiral anomaly-induced planar Hall effect, and is still visible at $T = 300$ K. Such a strong planar Hall effect provides another transport signature of chiral anomaly, and would ignite further research interest on chiral charge pumping in Dirac/Weyl semimetal systems.

# References

1. S. L. Adler, Phys. Rev. **177**, 2426 (1969).

2. J. S. Bell, R. Jackiw, Il Nuovo Cimento. A **60**, 47 (1969).

3. P. Hosur, X. L. Qi, C. R. Phys. **14**, 857 (2013).

4. D. T. Son, B. Z. Spivak, Phys. Rev. B **88**, 104412 (2013).

5. A. A. Burkov, Phys. Rev. Lett. **113**, 247203 (2014).

6. S. A. Parameswaran, T. Grover, D. A. Abanin, D. A. Pesin, and A. Vishwanath, Phys. Rev. X **4**, 031035 (2014).

7. A. A. Burkov, Phys. Rev. B **91**, 245157 (2015).

8. A. A. Burkov, Phys. Rev. B **96**, 041110(R) (2017).

9. S. Nandy, G. Sharma, A. Taraphder, and S. Tewari, arXiv: 1705.09308.

10. J. Xiong, S. K. Kushwaha, T. Liang, J. W. Krizan, M. Hirschberger, W. Wang, R. J. Cava, and N. P. Ong, Science **350**, 413 (2015).

11. C.-Z. Li, L.-X. Wang, H. Liu, J. Wang, Z.-M. Liao, and D.-P. Yu, Nat. Commun. **6**, 10137 (2015).

12. H. Li, H. He, H.-Z. Lu, H. Zhang, H. Liu, R. Ma, Z. Fan, S.-Q. Shen, and J. Wang, Nat. Commun. **7**, 10301(2015).

13. Q. Li, D. E. Kharzeev, C. Zhang, Y. Huang, I. Pletikosić, A. V. Fedorov, R. D. Zhong, J. A. Schneeloch, G. D. Gu, and T. Valla, Nat. Phys. **12**, 550-554 (2016).

14. G. L. Zheng, J. W. Lu, X. D. Zhu, W. Ning, Y. Y. Han, H. W. Zhang, J. L. Zhang, C. Y. Xi, J. Y. Yang, H. F. Du, K. Yang, Y. H. Zhang, and M. L. Tian,

Phys. Rev. B **93**, 115414 (2016).

15. T. Liang, Q. Gibson, M. Liu, W. Wang, R. J. Cava, and N. P. Ong, arXiv: 1612.06972.

16. M. Hirschberger, S. Kushwaha, Z. Wang, Q. Gibson, S. Liang, C. A. Belvin, B. A. Bernevig, R. J. Cava and N. P. Ong, Nat. Mater. **15**, 1161 (2016).

17. X. C. Huang, L. X. Zhao, Y. J. Long, P. P. Wang, D. Chen, Z. H. Yang, H. Liang, M. Q. Xue, H. M. Weng, Z. Fang, X. Dai, and G. F. Chen, Phys. Rev. X **5**, 031023 (2015).

18. C.-L. Zhang, S.-Y. Xu, I. Belopolski, Z. Yuan, Z. Lin, B. Tong, G. Bian, N. Alidoust, C.-C. Lee, S.-M. Huang, T.-R. Chang, G. Chang, C.-H. Hsu, H.-T. Jeng, M. Neupane, D. S. Sanchez, H. Zheng, J. Wang, H. Lin, C. Zhang, H.-Z. Lu, S.-Q. Shen, T. Neupert, M. Z. Hasan, and S. Jia, Nat. Commun. **7**, 10735 (2016).

19. A. C. Niemann, J. Gooth, S-C. Wu, S. Bäßler, P. Sergelius, R. Hühne, B. Rellinghaus, C. Shekhar, V. Süß, M. Schmidt, C. Felser, B. Yan, and K. Nielsch Sci. Rep. **7**, 43394 (2017).

20. P. Goswami, J. H. Pixley, and S. D. Sarma, Phys. Rev. B **92**, 075205 (2015).

21. F. Arnold, C. Shekhar, S.-C. Wu, Y. Sun, R. D. d. Reis, N. Kumar, M. Naumann, M. O. Ajeesh, M. Schmidt, A. G. Grushin, J. H. Bardarson, M. Baenitz, D. Sokolov, H. Borrmann, M. Nicklas, C. Felser, E. Hassinger, and B. Yan, Nat. Commun. **7**, 11615 (2016).

22. R. D. d. Reis, M. O. Ajeesh, N. Kumar, F. Arnold, C. Shekhar, M. Naumann, M.

Schmidt, M. Nicklas, and E. Hassinger, New J. Phys. **18**, 085006 (2016).

23. T. Schumann, M. Goyal, D. A. Kealhofer, and S. Stemmer, Phys. Rev. B **95**, 241113(R) (2017).

24. A. A. Taskin, H. F. Legg, F. Yang, S. Sasaki, Y. Kanai, K. Matsumoto, A. Rosch, and Y. Ando, arXiv: 1703.03406.

25. V. D. Ky, Zh. eksper. teor. Fiz. **50**, 1218 (1966).

26. H. X. Tang, R. K. Kawakami, D. D. Awschalom, and M. L. Roukes, Phys. Rev. Lett. **90**, 107201 (2003).

## Acknowledgments

This work was supported by the Natural Science Foundation of China (Grant No.11374302, U1432251).

# Figure Captions

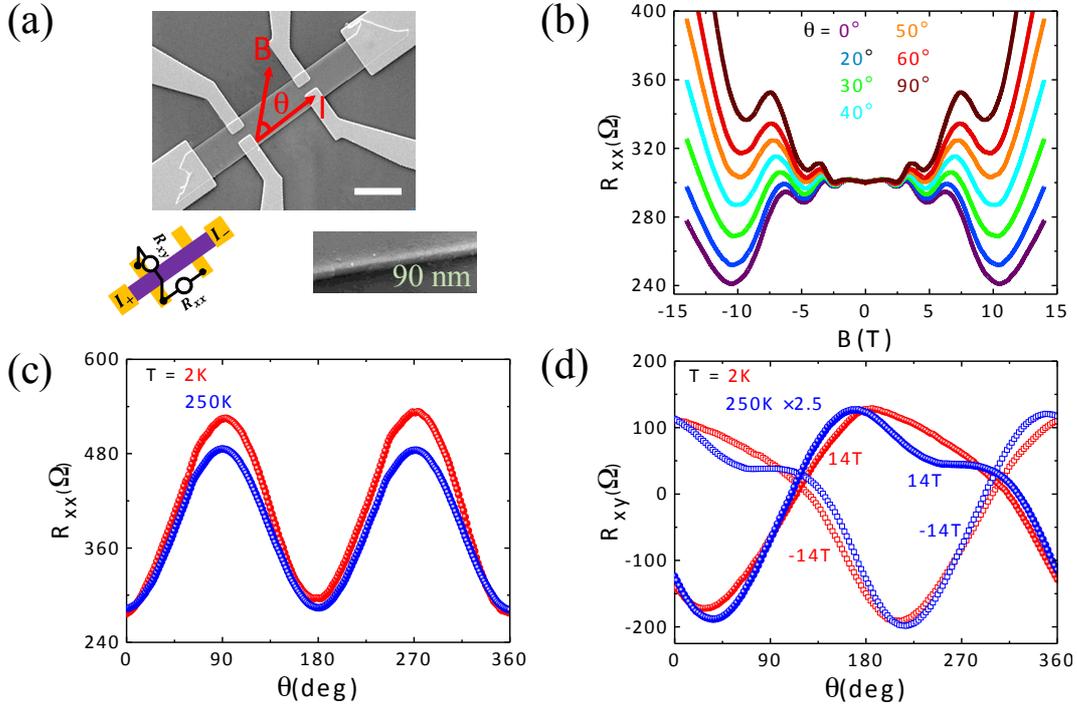

FIG. 1 (Color online). (a) The SEM image of a $Cd_3As_2$-based quantum device. Scale bar: $10\ \mu m$. Lower left inset: a schematic illustration of the measurement configuration. Lower right inset: The thickness of the sample determined by the $52°$-tilted SEM is about $90\ nm$. (b) The planar longitudinal resistance ($R_{xx}$) measured at 2 K with an in-plane magnetic field tilted from $\theta = 0°$ to $90°$. (c) Angular dependence of $R_{xx}$ with $|B| = 14$ T and $T = 2$ K and 250 K, respectively. (d) Shows the planar transverse magnetoresistance $R_{xy}$ with $B = \pm 14$ T, $T = 2$ K and 250 K, respectively.

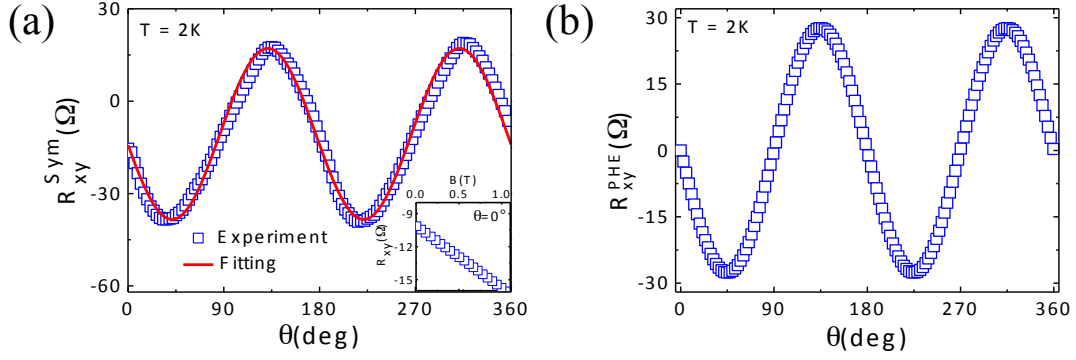

FIG. 2 (Color online). (a) The symmetrized planar TMR $R_{xy}^{sym}$ at $T = 2$ K and $|B| = 14$ T. Blue squares are the experimental data while red curve is the fitting curve to the formula (4). Inset: the magnetic field dependence of $R_{xy}$ at $\theta = 0°$, a non-zero transverse resistance of about $-10\ \Omega$ can be identified at $B = 0$ T, which comes from the small longitudinal resistance component due to the Hall bar misalignment. (b) The intrinsic planar Hall signals at $|B| = 14$ T after smearing out the misalignments. The obtained PHE exhibits a $-\sin\theta\cos\theta$ tendency.

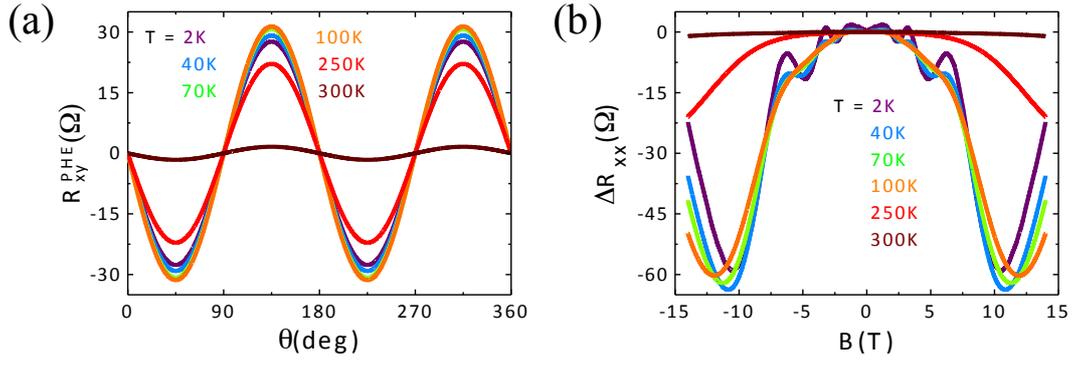

FIG. 3 (Color online). (a) The obtained PHE versus $\theta$ under various temperatures $T$. (b) The temperature-dependent negative LMR components, $\Delta R_{xx} = R(B) - R_0(B = 0)$. Both the PHE and the negative LMR decay synchronously as increasing the temperature, indicating the same origin of both effects.

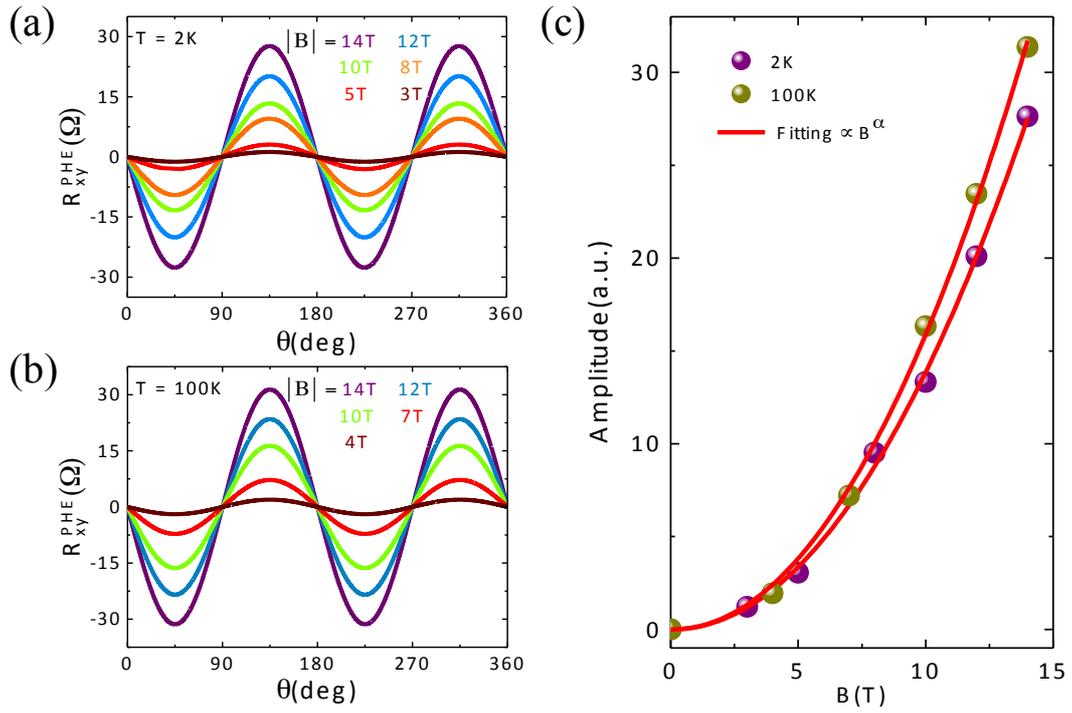

FIG. 4 (Color online). (a), (b) The field-dependent PHE at $T = 2$ K and 100 K, respectively. (c) The fittings of the amplitudes of PHE (dots) to the function $B^\alpha$. The amplitudes of PHE exhibit a $B^2$ tendency and they are not saturated even with B up to 14 T, as indicated by the red curves.